\documentclass{elsart} 
\journal{Physica D}
\date{4 March 2002}

\usepackage{amsmath}
\usepackage{amsfonts}

\usepackage{dcolumn}
\newcolumntype{d}{D{.}{.}{-1}}

\let\rem\undef

\usepackage[mediumspace,squaren,textstyle]{SIunits}

\usepackage{subfigure}

\usepackage[dvips]{graphicx}

\graphicspath{{:Figures:}{Figures/}}

\newcommand{\Barrier}[1]{\ensuremath{W_{#1}}}
\newcommand{\ChargeDensity}{\ensuremath{\rho}}
\newcommand{\Chemical}[1]{\ensuremath{\mu_{#1}}}
\newcommand{\Components}{\ensuremath{N}}
\newcommand{\Concentration}[1]{\ensuremath{C_{#1}}}
\newcommand{\Curvature}{\ensuremath{K}}
\newcommand{\Galvani}{\ensuremath{\Delta\Potential^\text{Ref}}}
\newcommand{\Dielectric}{\ensuremath{\epsilon}}
\newcommand{\DoubleWell}{\ensuremath{g}}
\newcommand{\Electrochemical}[1]{\ensuremath{\bar{\mu}_{#1}}}
\newcommand{\Electrode}{\ensuremath{S}}
\newcommand{\Electrolyte}{\ensuremath{L}}
\newcommand{\Faraday}{\ensuremath{\mathcal{F}}}
\newcommand{\Fraction}[1]{\ensuremath{X_{#1}}}
\newcommand{\Gas}{\ensuremath{R}}
\newcommand{\Gradient}[1]{\ensuremath{\kappa_{#1}}}
\newcommand{\Helmholtz}{\ensuremath{F}}
\newcommand{\HelmholtzPerVol}{\ensuremath{f_{\Volume}}}
\newcommand{\Interpolate}{\ensuremath{p}}
\newcommand{\MolarVolume}{\ensuremath{\Volume_{m}}}
\newcommand{\PartialMolarVolume}[1]{\ensuremath{\bar{\Volume}_{#1}}}
\newcommand{\Phase}{\ensuremath{\varphi}}
\newcommand{\Position}{\ensuremath{x}}
\newcommand{\Potential}{\ensuremath{\psi}}
\newcommand{\Scientific}[2]{\ensuremath{#1\times\power{10}{#2}}}
\newcommand{\Solvent}{\ensuremath{s}}  
\newcommand{\Standard}[2][]{\ensuremath{\Chemical{#2}^{\circ#1}}}
\newcommand{\Substitutional}{\ensuremath{S}}
\newcommand{\SurfaceCharge}{\ensuremath{\sigma}}
\newcommand{\SurfaceEnergy}{\ensuremath{\gamma}}
\newcommand{\Temperature}{\ensuremath{T}}
\newcommand{\Valence}[1]{\ensuremath{z_{#1}}}
\newcommand{\Volume}{\ensuremath{V}}

\newcommand{\Cupric}{\ensuremath{\text{Cu}^{+2}}}
\newcommand{\Sulfate}{\ensuremath{\text{SO}_{4}^{-2}}}
\newcommand{\Copper}{\text{Cu}}
\newcommand{\CopperSulfate}{\ensuremath{\text{CuSO}_{4}}}
\newcommand{\Electron}{\ensuremath{\text{e}^{-}}}
\newcommand{\Water}{\ensuremath{\text{H}_{2}\text{O}}}

\begin{document}
    
\begin{frontmatter}
    
\title{Model of Electrochemical ``Double Layer'' Using the Phase Field Method}
\author{J. E. Guyer\corauthref{cor}},
\corauth[cor]{Corresponding author.}
\ead{guyer@nist.gov}
\author{W. J. Boettinger},
\author{J. A. Warren}
\address{Metallurgy Division}

\author{G. B. McFadden}
\address{Mathematical and Computational Sciences Division}

\address{National Institute of Standards and Technology, 
Gaithersburg, MD 20899, USA}
    
\begin{abstract}

We present the first application of phase field modeling to
electrochemistry.  A free energy functional that includes the
electrostatic effect of charged particles leads to rich interactions
between concentration, electrostatic potential, and phase stability. 
The present model, explored for the equilibrium planar interface,
properly predicts the charge separation associated with the
equilibrium double layer at the electrochemical interface and its
extent in the electrolyte as a function of electrolyte concentration,
as well as the form expected of surface energy vs.  potential
(``electrocapillary'') curves.
\end{abstract}

\end{frontmatter}

\section{Introduction}

The phase field technique has previously been applied to the time
evolution of complex dendritic, eutectic, and peritectic
solidification morphologies \cite{Boettinger:1999,Drolet:2000}.  The
present work was motivated by the mathematical analogy between the
governing equations of solidification dynamics and electroplating
dynamics.  For example, the solid-liquid interface is analogous to the
electrode-electrolyte interface.  The various overpotentials of
electrochemistry have analogies with the supercoolings of alloy
solidification: diffusional (constitutional), curvature, and interface
attachment.  Dendrites can form during solidification and during
electroplating.  Nonetheless, we find significant differences between
the two systems.

Electrochemistry has been chosen by the microelectronics industry as
the deposition method for ``copper damascene'' interconnects on
microchips because it is capable of filling trenches and vias
\cite{IBM:1998,NIST:damascene:2001} without ``pinch-off'' void
formation problems associated with chemical or physical vapor
deposition.  Because plating into such small features naturally
involves large curvatures of the growing interface and large electric
field gradients, a rigorous mathematical solution of the governing
equations is warranted.

Modeling the evolution of a sharp interface, particularly when the
topology is changing during growth, is a challenging problem because
the boundary conditions at that interface are dependent on the shape
and location of the interface.  The phase field method avoids these
difficulties by introducing a new phase variable and an appropriate
governing equation for that variable.  The governing equations for the
system can then be solved on a uniform grid, with the interface
position as one of the results of the solution.  A simple analytical
electrochemical cell consists of a working electrode, an electrolyte,
and a reference electrode.  In the case of electroplating, there will
also be a counter electrode.  The phase field variable in this work
describes the transition between the working electrode and the
electrolyte.

\subsection{Length scales in electrochemistry}

\begin{figure}[tbp]
    \vspace*{-0.5in}
    \centering
    \centerline{\includegraphics[width=5in]{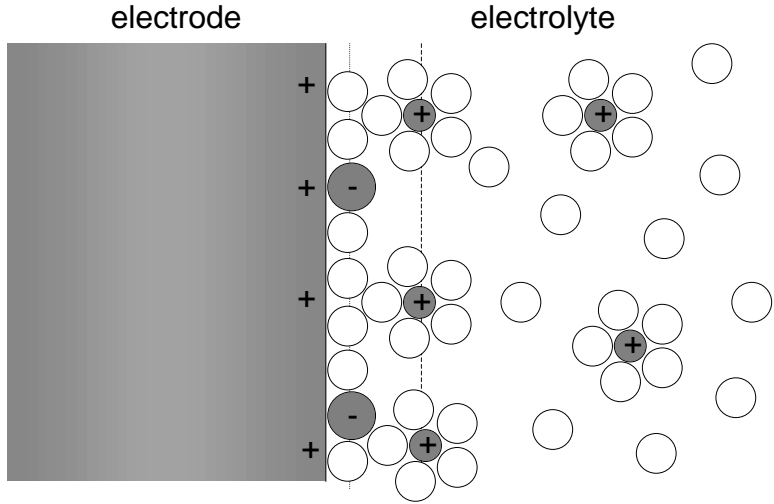}}
    \vspace*{-0.35in}
    \caption{Schematic of the electrode-electrolyte interface.}
    \protect\label{fig:DoubleLayer}
\end{figure}

There are three significant length scales in 
electrochemistry:
\begin{enumerate}
    \item the thickness of the electrode-electrolyte interface,

    \item the charge separation (of the capacitive double layer) and
    voltage decay distance, which are related to the concentrations of
    charged species and the dielectric constant,

    \item and the long range concentration decay length due to 
    diffusion and convection in the electrolyte (related to the ratio 
    of diffusivity to interfacial velocity).
\end{enumerate}
In traditional modeling of the equilibrium electrochemical interface,
the electrode-electrolyte interface is assumed to be sharp, as shown
in Figure~\ref{fig:DoubleLayer}.
The charge separation distance is considered diffuse in the
electrolyte only, with any excess charge on the electrode residing
entirely at the metal surface.  In the electrolyte, there may be some
specific adsorption of ions on the interface, and a more diffuse
distribution of ions further away.  Some ions may be complexed with
solvent molecules, giving them a larger effective size and limiting
their approach to the electrode surface (we do not consider solvation
in this work).  The long range concentration decay length is absent
for an equilibrium interface.  This equilibrium regime is
characteristic of models such as that of Gouy and Chapman.  By
requiring that the ions in solution must simultaneously satisfy a
Boltzmann distribution and Poisson's equation, a description of the
charge distribution is obtained, with predictions for the surface
energy, charge separation, and adsorption.

In another limit of traditional modeling, primarily used for dynamics,
the electrode-electrolyte interface is assumed sharp and the charge
separation is a dipole layer.  The long range concentration decay
length is modeled by transport in a stagnant fluid boundary layer.  In
equilibrium, the concentration and electrostatic potential jump across
the interface are related by the Nernst equation
\begin{equation}
    \Potential^\text{eq} = \Potential^\circ
    +\frac{\Gas\Temperature}{\Valence{}\Faraday}
    \ln\frac{\Concentration{O}}{\Concentration{R}}
    \label{eq:Nernst}
\end{equation}
where \( \Potential^\text{eq} \) is the equilibrium electrostatic
potential jump, \( \Potential^{\circ} \) is the standard potential,
\Gas\ is the molar gas constant, \Faraday\ is Faraday's constant,
\Temperature\ is the temperature, \Concentration{O} is the
concentration of the electroactive species in the oxidized state,
\Concentration{R} is the concentration in the reduced state,
and \Valence{} is the number of electrons transferred in the redox
reaction.  A term \( -\MolarVolume\SurfaceEnergy\Curvature /
\Valence{}\Faraday \) is added to the right hand side of
Eq.~\eqref{eq:Nernst} to treat curved interfaces, where \MolarVolume\
is the molar volume, \SurfaceEnergy\ is the surface energy, and
\Curvature\ is the curvature.  In dynamic conditions, a chemical
reaction rate description (such as the Butler-Volmer relation) is used
to describe the relationship between potential jump and current.

In the present phase field model, all three length scales are assumed
to be important (albeit different) and all are modeled
together with the same equations in the metal and electrolyte.  This
simplifies treatment of complex geometries and permits proper
treatment of adsorption and its effect on interface kinetics.  As
usual for phase field models, though, the equations can be very stiff,
requiring significant computational resources.

\subsection{Traditional double-layer theory (Gouy-Chapman)}

The best known sharp-interface model of the electrochemical interface
is that of Gouy and Chapman \cite{Bard:2nd,Grahame:1947}.  In this
model (and its variants), the electrode is not considered, the
distribution of ions in the electrolyte is assumed to follow a
Boltzmann distribution
\begin{equation}
    \Concentration{j} = \Concentration{j}^{\infty}\exp\left[
    	\frac{-\Valence{j}\Faraday\left(\Potential-\Potential_{\infty}\right)}
	{\Gas\Temperature}
    \right]
    \label{eq:Boltzmann}
\end{equation}
and to satisfy Poisson's equation
\begin{equation}
    \ChargeDensity=\sum_{\Components}\Concentration{j}\Valence{j}\Faraday
    = -\Dielectric\frac{d^2 \Potential}{d \Position^2}.
    \label{eq:Poisson}
\end{equation}
\Concentration{j} is the concentration of species \( j \), \(
\Concentration{j}^{\infty} \) is the concentration of species \( j \)
in the bulk electrolyte, \Valence{j} is the valence of species \( j
\), \ChargeDensity\ is the charge density, \Dielectric\ is the
permittivity of the electrolyte, and the electrostatic potential field
\( \Potential = \Potential_{\infty} \) in the far field.  In a
\Valence{}:\Valence{} electrolyte (cation and anion have equal and
opposite valence), these equations reduce to
\begin{equation}
    \frac{d \Potential}{d \Position} = -\left(
    	\frac{8\Gas\Temperature\Concentration{}^{\infty}}{\Dielectric}
    \right)^{1/2}
    \sinh\left[
	\frac{\Valence{}\Faraday\left(\Potential-\Potential_{\infty}\right)}
	{2\Gas\Temperature}
    \right],
    \label{eq:GouyChapman}
\end{equation}
which has the solution
\begin{equation}
    \frac{\tanh\left[
	\Valence{}\Faraday\left(\Potential-\Potential_{\infty}\right)
	/4\Gas\Temperature
    \right]}
    {\tanh\left[
	\Valence{}\Faraday\left(\Potential_{0}-\Potential_{\infty}\right)
	/4\Gas\Temperature
    \right]}
    =\exp\left(-\Position/l_{\Potential}\right).
    \label{eq:GouyChapmanSolution}
\end{equation}
\( \Potential_0 \) is the electrostatic potential in the electrolyte
at the electrode surface (\( \Position = 0 \)) and \( l_{\Potential} =
[\Dielectric\Gas\Temperature/(2\Concentration{}^{\infty}\Valence{}^2\Faraday^2)]^{1/2}
\) is the voltage/concentration decay length (the same as the Debye
length).

Using this model, the surface energy, surface charge, differential 
capacitance, adsorption and other interfacial parameters can be 
related to the voltage across the interface. The surface charge on 
the electrode \( \SurfaceCharge^{\Electrode} \) is described by
\begin{equation}
    \SurfaceCharge^{\Electrode} = 
    -\left(
	\frac{\partial\SurfaceEnergy}{\partial\Potential_{0}}
    \right)_{\Chemical{i}}
    \label{eq:SurfaceChargeGC}
\end{equation}
where \SurfaceEnergy\ is the surface energy, and subscript
\Chemical{i} denotes that the partial derivative is taken at constant
chemical potential, \emph{i.e.}, at constant activities (or
concentrations) of all species, as approximated on an inert mercury
electrode.  A plot of surface energy as a function of \(
\Potential_{0} \) is known as an ``electrocapillary curve'' and from
Eq.~\eqref{eq:SurfaceChargeGC} we see that its maximum corresponds to
the so-called ``potential of zero charge'', where there is no net
surface charge on the electrode (and due to overall charge balance,
none in the electrolyte).

\section{Phase Field Model}

\subsection{Components}

The phase field approach requires a set of equations that can describe
the evolution of conditions at every point in the system.  Thus we
must identify the chemical components that describe the electrode and
electrolyte simultaneously.  As our model problem, we consider a
\Copper\ metal electrode in contact with a \CopperSulfate\ aqueous
electrolyte.  We choose our electrode to be a solid solution of
\Cupric\ and interstitial \Electron.  We choose our electrolyte to be
an aqueous solution of \Water, \Cupric, and \Sulfate.  We assume that
the partial molar volume of \Electron\ is zero and that the partial
molar volumes of \Cupric, \Water, and \Sulfate\ are all the same, such
that
\begin{equation}
    \MolarVolume = \sum^{\Components} \PartialMolarVolume{j} \Fraction{j}
    = \PartialMolarVolume{\Substitutional} \sum^{\Substitutional} 
    \Fraction{j}
\end{equation}
and    
\begin{equation}
    \sum_{\Substitutional} \Concentration{j} = 
    \frac{1}{\PartialMolarVolume{\Substitutional}},
    \label{eq:substitutionalConstraint}
\end{equation}
where \MolarVolume\ is the molar volume, \( \Fraction{j} =
\Concentration{j} \MolarVolume \) is the mole fraction of component \(
j \) with partial molar volume \PartialMolarVolume{j}, \Components\ is
the set of all components, and \Substitutional\ is the set of
substitutional components with partial molar volume
\PartialMolarVolume{\Substitutional}.  This treatment of \Electron\ as
interstitials allows the motion of \Electron\ to decouple from the
other species.  The simplification that all substitutional species
have the same, constant partial molar volume allows us to consider
diffusion separately from deformation.

\subsection{Free energy}

The free energy of this two phase system of \Components\ charged species is 
given by
\begin{equation}
    \Helmholtz\left(
        \Phase, \Concentration{1},\ldots ,\Concentration{\Components}, 
        \Potential
    \right)
    = \int_{\Volume}\left[
    	\HelmholtzPerVol\left(
            \Phase,\Concentration{1},\ldots ,\Concentration{\Components}
        \right)
        +\frac{1}{2}\ChargeDensity\Potential
        +\frac{\Gradient{\Phase}}{2}\left|\nabla\Phase\right|^2
    \right] \d\Volume
    \label{eq:Helmholtz}
\end{equation}
where \Helmholtz\ is the total Helmholtz free energy,
\HelmholtzPerVol\ is the Helmholtz free energy per unit volume,
\Phase\ is the phase field variable, and \( \Gradient{\Phase} \) is
the phase field gradient energy coefficient.  \( \Phase = 1 \) denotes
the electrode and \( \Phase = 0 \) denotes the electrolyte.  The first
term in the integral of Eq.~\eqref{eq:Helmholtz} describes the
chemical energy, the second term describes the electrostatic energy,
and the third term describes the interfacial energy of the phase
field.

\subsection{Equilibrium}

We compose a Lagrangian from Eq.~\eqref{eq:Helmholtz} by requiring
that mass of each species must be conserved, and that Poisson's equation
\eqref{eq:Poisson} and the substitutional constraint
Eq.~\eqref{eq:substitutionalConstraint} must be satisfied everywhere
in the system.  By taking the virtual variations of this Lagrangian,
we find that the governing equations in equilibrium are
\begin{equation}
    \Electrochemical{j} - 
    \frac{\PartialMolarVolume{j}}{\PartialMolarVolume{\Substitutional}}
    \Electrochemical{\Solvent}
    = \text{constant},
    \label{eq:electrochemical}
\end{equation}
\begin{equation}
    0 
    = \frac{\partial\HelmholtzPerVol}{\partial\Phase}
    - \Gradient{\Phase}\nabla^2\Phase
    - \frac{\Dielectric'\left(\Phase\right)}{2}\left(\nabla\Potential\right)^2,
    \label{eq:phaseField}
\end{equation}
and Poisson's equation \eqref{eq:Poisson}.  
The electrochemical potential of species \( j \) is
\begin{equation}
    \Electrochemical{j} = 
        \frac{\partial\HelmholtzPerVol}{\partial\Concentration{j}}
        + \Faraday\Valence{j}\Potential.
\end{equation}
The subscript \Solvent\ denotes the
solvent, chosen arbitrarily from the set \Substitutional. The complete 
derivation of these equations is given in Ref.~\cite{ElPhF:2002}.

\subsection{Choice of thermodynamics}

\begin{figure}[tbp]
    \centerline{\includegraphics[width=3in]{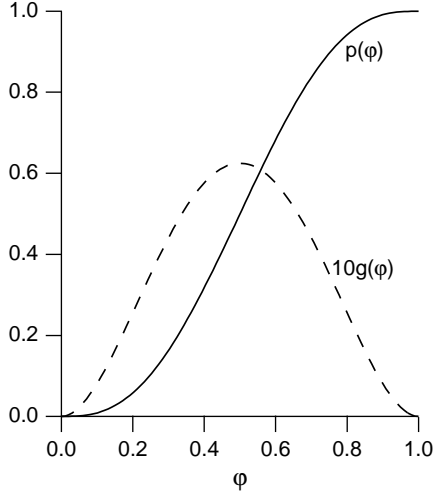}}
    \caption{Phase field double well and interpolation functions.}
    \protect\label{fig:g&p}
\end{figure}

To simplify our model as much as possible, we choose to describe the 
chemical Helmholtz free energy by ideal solution thermodynamics
\begin{equation}
    \HelmholtzPerVol\left(\Phase,\Concentration{j}\right)
    = \sum_{\Components}\Concentration{j}\left\{
    	\Standard[\Electrolyte]{j}
    	+ \Delta\Standard{j}\Interpolate\left(\Phase\right)
        + \Gas\Temperature\ln\Concentration{j}\MolarVolume
        + \Barrier{j}\DoubleWell\left(\Phase\right)
    \right\}
    \label{eq:IdealHelmholtz}
\end{equation}
such that
\begin{equation}
    \Electrochemical{j}
    = \Standard[\Electrolyte]{j}
    + \Delta\Standard{j}\Interpolate\left(\Phase\right)
    + \Gas\Temperature\ln\Concentration{j}\MolarVolume
    + \Valence{j}\Faraday\Potential
    + \Barrier{j}\DoubleWell\left(\Phase\right).
    \label{eq:IdealElectrochemical}
\end{equation}
\Standard[\Electrolyte]{j} is the standard chemical potential of
component \( j \) in the electrolyte, \( \Delta\Standard{j} \) is the
difference between the standard chemical potentials of component \( j \)
in the electrode and in the electrolyte, and \Barrier{j} is the
barrier height of the phase-field double well for component \( j \). 
Other choices are possible, but we select
\begin{equation}
    \Interpolate\left(\Phase\right)
    = \Phase^3\left(6\Phase^2-15\Phase+10\right)
    \label{eq:Interpolation}
\end{equation}
for the phase field interpolation function \Interpolate, which is a
smoothed step function, and
\begin{equation}
    \DoubleWell\left(\Phase\right)
    = \Phase^2\left(1-\Phase\right)^2
    \label{eq:DoubleWell}
\end{equation}
for the phase field double well function \DoubleWell.  These functions
are plotted in Figure~\ref{fig:g&p}.

\subsection{Choice of parameters}

If we consider only the values of concentration and potential far from
an equilibrium interface (\( \Phase = 0 \), \( \Phase = 1 \)), we can
make contact with the sharp interface picture.  Analysis of
Eqs.~\eqref{eq:electrochemical} and \eqref{eq:phaseField} yields the
fact that the far field values of the electrochemical potentials
\Electrochemical{j} are equal.  From
Eq.~\eqref{eq:IdealElectrochemical} (evaluated at \( \Phase = 0 \) and
\( \Phase = 1 \)), we see that, for all species,
\begin{equation}
    \Potential^{\Electrode} - \Potential^{\Electrolyte}
    = \frac{\Gas\Temperature}{\Valence{j}\Faraday}
        \ln\frac{\Fraction{j}^{\Electrolyte}}{\Fraction{j}^{\Electrode}}
        -\frac{\Delta\Standard{j}}{\Valence{j}\Faraday}.
    \label{eq:Nernst:Ours}
\end{equation}
These equations describe the relationship between the potential
difference and the concentration of each component in the two phases. 
Normally in electrochemistry, the electrode is considered to be pure
metal and the equations are only applied to the species involved in
electron transfer.  Evaluated in this situation,
Eq.~\eqref{eq:Nernst:Ours} is equivalent to the 
Nernst equation
\eqref{eq:Nernst}.

From Eq.~\eqref{eq:Nernst:Ours}, we see that the standard
chemical potential difference between the electrode and electrolyte \(
\Delta\Standard{j} \) can be broken into two terms
\begin{equation}
    \Delta\Standard{j} 
    = \Gas\Temperature\ln
    \frac{\Fraction{j}^{\Electrolyte, \text{Ref}}}
        {\Fraction{j}^{\Electrode, \text{Ref}}}
    - \Valence{j}\Faraday\Galvani.
    \label{eq:DeltaStandard}
\end{equation}
\( \Galvani = \left(\Potential^{\Electrode} -
\Potential^{\Electrolyte}\right)^\text{Ref} \) and superscript \(
\text{Ref} \) denotes the reference state for the standard potentials. 
The chosen value of the parameter \Galvani\ can be shown to have a
direct relationship with the surface energy of the electrochemical
interface, the charge on the metal, and the deviation of the interface
from the point of zero charge.  As such, it is a materials parameter
that is fixed for each metal/electrolyte system, even as the actual
concentrations are changed.  \Galvani\ is equivalent to the
``Galvani'' potential of the electrode-electrolyte interface, and is
related to the difference between the work functions of the electrode
and electrolyte.

We choose as our reference state a metallic copper electrode in 
contact with a \unit{1}{\mole\per\liter} \CopperSulfate\ aqueous 
electrolyte, as described in Table~\ref{tab:Parameters:Thermodynamic}.
\begin{table}[b]
	\centering
	\caption{Values of reference concentrations.}
	\begin{tabular}{cdddd}
            \\
        \hline
            \vphantom{\( \frac{\frac{A}{B}}{\frac{B}{C}} \)}
            & \multicolumn{1}{c}{$\Fraction{j}^{\Electrolyte, \text{Ref}}$}
            & \multicolumn{1}{c}{
                \unit{\Concentration{j}^{\Electrolyte, \text{Ref}}}
                    {\per(\mole\per\meter\cubed)}
                }
            & \multicolumn{1}{c}{$\Fraction{j}^{\Electrode, \text{Ref}}$}
            & \multicolumn{1}{c}{
                \unit{\Concentration{j}^{\Electrode, \text{Ref}}}
                    {\per(\mole\per\meter\cubed)}
                }
        \\

        \hline

        $ {\Cupric} $
        & 0.0180
        & 1000
        & 0.333
        & 55600
        \\
        
        $ {\Sulfate} $
        & 0.0180
        & 1000
        & \Scientific{1.0}{-6}
        & 0.167
        \\
        
        $ {\Electron} $
        & \Scientific{1.0}{-6}
        & 0.0556
        & 0.667
        & 111000
        \\
        
        $ {\Water} $
        & 0.964
        & 53600
        & \Scientific{9.99}{-7}
        & 0.167
        \\
        
        \hline
        
    \end{tabular}
	\protect\label{tab:Parameters:Thermodynamic}
\end{table}
In order to perform equilibrium numerical simulations, some trace of
\Water\ and \Sulfate\ must be permitted in the electrode and some
trace of \Electron\ must be permitted in the electrolyte.  We
arbitrarily choose a mole fraction of \power{10}{-6} for these trace
components in the reference state.
The remaining parameters are \( \Barrier{\Electron} = 
\unit{0}{\joule\per\mole} \), all other \( \Barrier{j} = 
\unit{\Scientific{3.6}{5}}{\joule\per\mole} \), \( 
\Gradient{\Phase} = \unit{\Scientific{3.6}{-11}}{\joule\per\meter} \), 
\( \PartialMolarVolume{\Solvent} = 
\unit{\Scientific{1.8}{-5}}{\meter\cubed\per\mole} \), and \( 
\Dielectric = \unit{\Scientific{6.95}{-10}}{\farad\per\meter} \).

\begin{figure}[tbp]
    \centering
    \subfigure[Electrolyte]{\label{fig:Nernst:Electrolyte}%
        \includegraphics[width=2.5in]{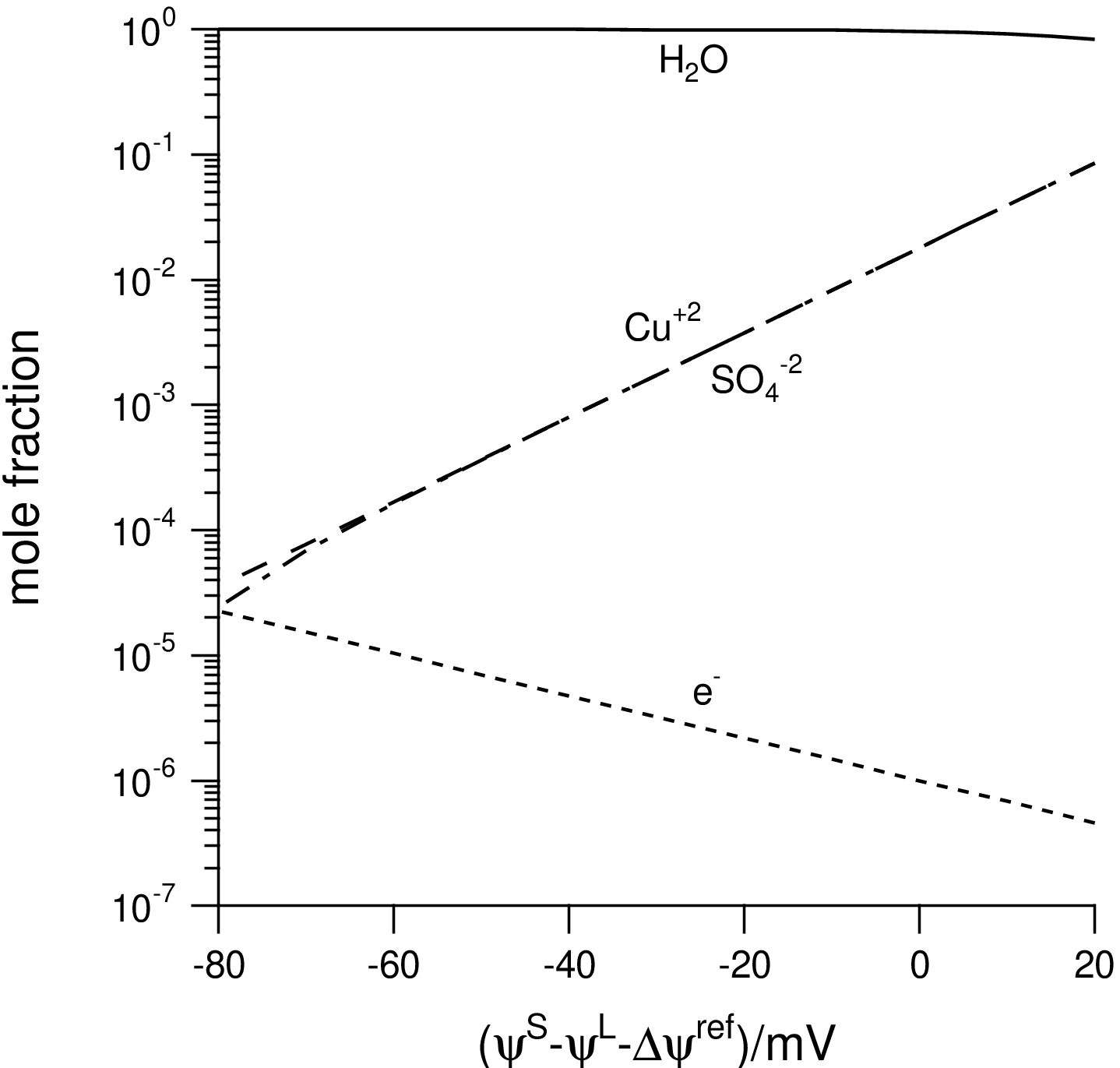}}
    \subfigure[Electrode]{\label{fig:Nernst:Electrode}%
        \includegraphics[width=2.5in]{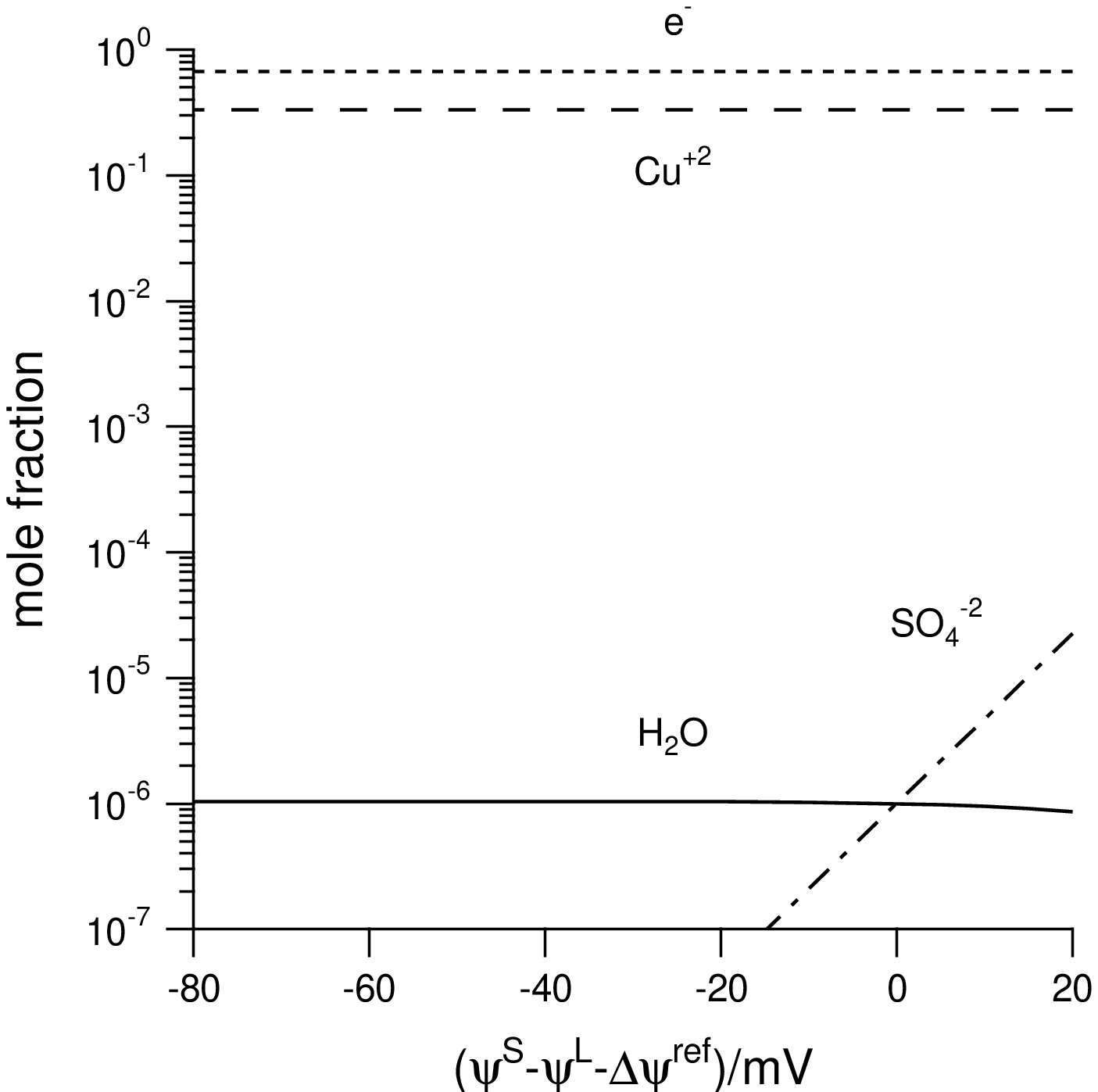}}%
    \caption{Mole fraction as a function of applied potential for the 
    electrode-electrolyte equilibrium. Note that the slope of 
    \Sulfate\ in the electrode is twice that in the electrolyte.}
    \label{fig:Nernst}
\end{figure}

Fixing the reference concentrations, we can examine the consequences
of varying the potential difference around \Galvani.  We
simultaneously solve Eq.~\eqref{eq:Nernst:Ours} for all four
components, subject to charge neutrality and \( \sum^{\Components}
\Fraction{j} = 1 \) in each phase.  The resulting mole fractions of
the electrode and electrolyte are shown in Figure~\ref{fig:Nernst}.
We can see that, over the entire range of \Galvani, the
electrode remains copper metal, with trace amounts of \Water\ and
\Sulfate.  Over most of the displayed range, the electrolyte is
predominantly water, with \Cupric\ and \Sulfate\ obeying Nernstian
behavior.  

The behavior of the trace amount of \Sulfate\ in the electrode
warrants discussion.  If we assume the \Sulfate\ concentration of the
electrode is fixed, the Nernst relationship for \Sulfate\ indicates
that its concentration in the electrolyte should move in the opposite
direction from \Cupric\ in the electrolyte, for a given change in \(
\Potential^{\Electrode} - \Potential^{\Electrolyte} \), because of the
sign change on \Valence{j}.  The requirement of bulk charge neutrality
in the electrolyte prevents this from occurring, however.  If the
concentration of \Cupric\ in the electrolyte increases by an order of
magnitude, charge neutrality requires that the concentration of
\Sulfate\ in the electrolyte increases by the same amount (neglecting
\Electron).  In order for Eq.~\eqref{eq:Nernst:Ours} to be satisfied
for both \Cupric\ and \Sulfate, \emph{i.e.}, for all species to be in
electrochemical equilibrium, the concentration of \Sulfate\ in the
electrode must then increase by \emph{two} orders of magnitude.

\begin{figure}[tbp]
    \centerline{\includegraphics[width=3in]{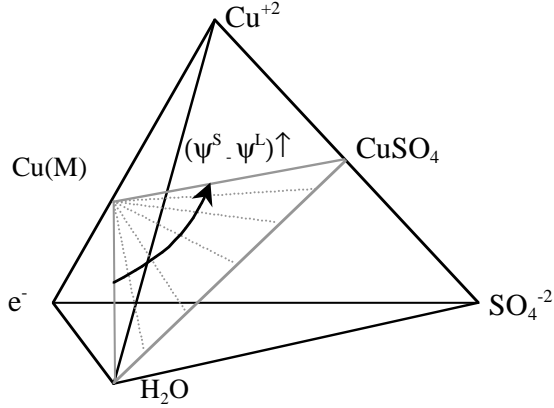}}
    \caption{Phase diagram and charge-neutral plane for 
    \Copper(M)--\CopperSulfate--\Water.}
    \protect\label{fig:Quaternary}
\end{figure}

Another way to see the equilibrium relationship between these phases 
is on the quaternary phase diagram in Figure~\ref{fig:Quaternary}. 
The vertices of the diagram are pure \Cupric, \Sulfate, \Water, and
\Electron.  The plane shows the zero charge combinations and the tie
lines between liquid and solid phases show the equilibria varying
between \Copper(M)--\Water\ at negative electrode potentials and
\Copper(M)--\CopperSulfate\ at positive electrode potentials.

\subsection{Equilibrium profiles}

\begin{figure}[tbp]
    \centerline{\includegraphics[width=5in]{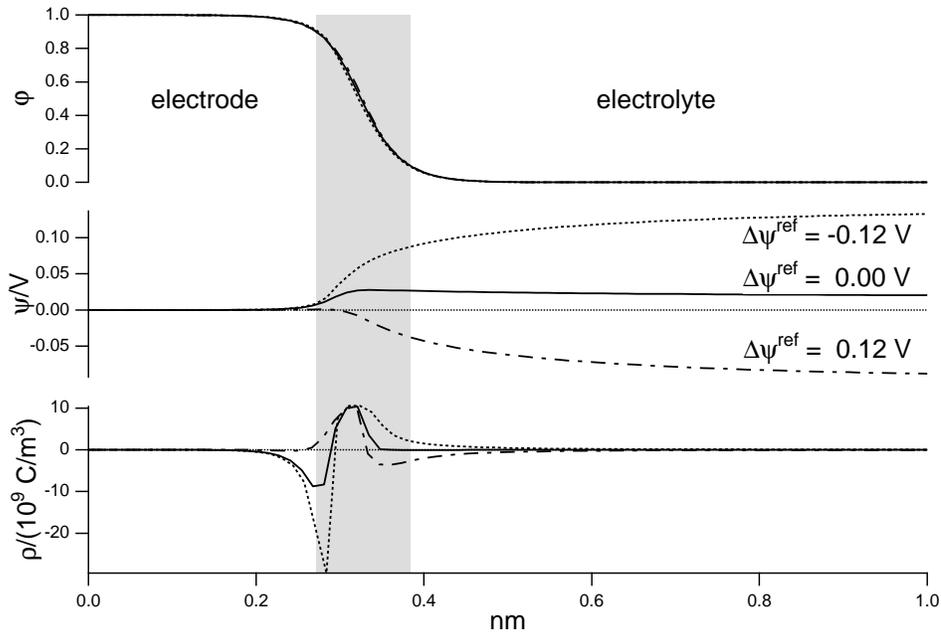}}
    \caption{Profiles through the 1D equilibrium interface for
    different values of \Galvani.  The shaded area denotes the
    interfacial region \( 0.1 < \Phase < 0.9 \).}
    \protect\label{fig:Profiles}
\end{figure}

\begin{figure}[tbp]
    \centerline{\includegraphics[width=5in]{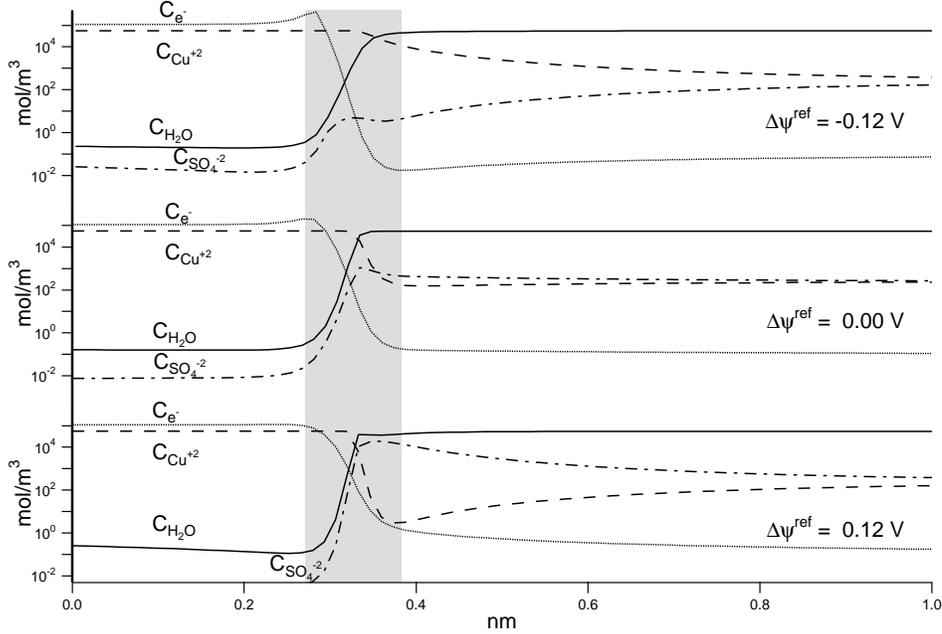}}
    \caption{Concentration profiles through the 1D equilibrium
    interface for different values of \Galvani.  The shaded area
    denotes the interfacial region \( 0.1 < \Phase < 0.9 \).}
    \protect\label{fig:ConcProfiles}
\end{figure}

Eqs.~\eqref{eq:Poisson}, \eqref{eq:electrochemical}, and
\eqref{eq:phaseField} are solved with finite difference techniques (to
second order in space) on a uniform grid.  The sharp, two-phase
initial condition was allowed to relax to equilibrium.  The numerical
methods are discussed in detail in Ref.~\cite{ElPhF:2002}.  The 1D
interface between electrode and electrolyte is shown in
Figures~\ref{fig:Profiles}
and \ref{fig:ConcProfiles}.  
The different curves are for fixed concentration and different values
of \Galvani.  We see that, as one might expect, the phase field
\Phase\ is not particularly sensitive to \Galvani, but the charge
distribution \ChargeDensity\ is.  All of the \Phase\ curves fall
within 2\% of a hyperbolic tangent profile, but the charge density
changes from a negatively charged electrode at \( \Galvani =
\unit{-0.12}{\volt} \) to a positively charged electrode at \(
\Galvani = \unit{+0.12}{\volt} \).  At \( \Galvani =
\unit{0}{\volt} \), we see that there is essentially zero charge in
the electrolyte, and the dipole charge in the electrode likewise sums
to a very small charge (note that a dipole layer does not necessarily
imply the presence of polar molecules \cite{Grahame:1947}).

The charge distributions in Figure~\ref{fig:Profiles} are a result of 
the ionic distributions in Figure~\ref{fig:ConcProfiles}. At \( \Galvani = 
\unit{-0.12}{\volt} \), there is a surplus of \Electron\ at the 
electrode interface. \Cupric\ remains at essentially the electrode 
density to some distance into the electrolyte, indicating an adsorbed 
layer of \Cupric\ (note the specific adsorption of \Sulfate\ in 
each case).

\begin{figure}[tbp]
    \centerline{\includegraphics[width=5in]{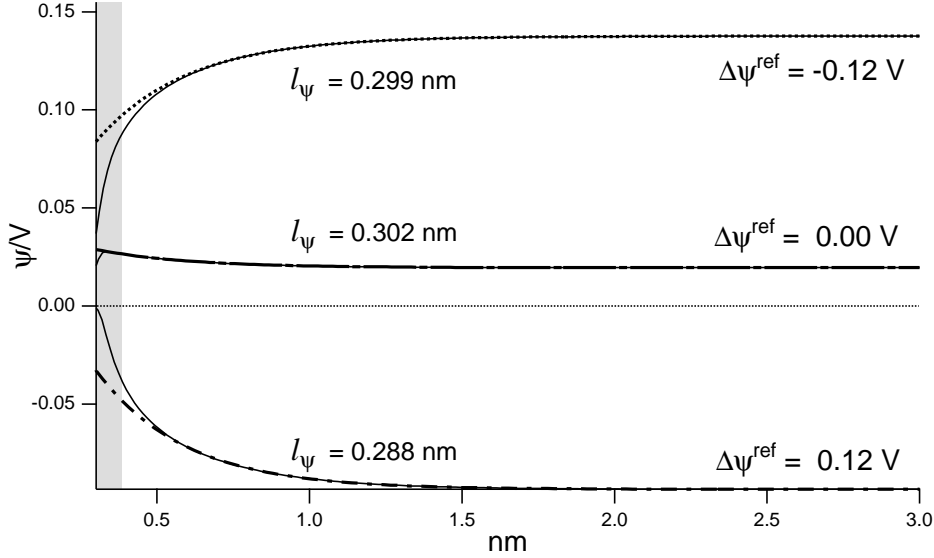}}
    \caption{Comparison of electrostatic potential profiles (light
    solid lines) with predictions of Gouy-Chapman model (heavy dashed
    lines). The shaded area denotes the interfacial region \( 0.1 < 
    \Phase < 0.9 \).}
    \protect\label{fig:PotentialProfiles}
\end{figure}

We examine the electrostatic potential profiles of Figure~\ref{fig:Profiles} 
in Figure~\ref{fig:PotentialProfiles}. 
We fit each of the curves to \( \Potential = \Potential_{\infty} +
\left(\Potential_{0} - \Potential_{\infty}\right)
\exp\left(-\Position/l_{\Potential}\right) \) and find excellent agreement for
the decay length with \( l_{\Potential} = \unit{0.304}{\nano\meter} \) predicted
by the Gouy-Chapman model.  The deviation between fit and model at the
left edge of the plot are because the Gouy-Chapman model assumes an
abrupt electrode-electrolyte interface and the phase field model
treats it as diffuse.

\subsection{Interface properties}

From a 1D analysis of Eq.~\eqref{eq:Helmholtz}, 
the equilibrium surface energy is described by
\begin{equation}
    \SurfaceEnergy = \int_{-\infty}^{\infty}\left[
    	\Gradient{\Phase}\left(\Phase'\right)^2
        -\Dielectric\left(\Potential'\right)^2
    \right]\,\d\Position
    \label{eq:SurfaceEnergy}
\end{equation}
where the prime denotes the derivative with respect to position.  
The surface charge on the electrode is related to the surface energy by
\begin{equation}
    -\frac{\partial\SurfaceEnergy}{\partial\Galvani}
    = \SurfaceCharge^{\Electrode}
    \equiv
    \int_{-\infty}^{\infty}
    	\Interpolate\left(\Phase\right)\ChargeDensity\,\d\Position
    \label{eq:SurfaceCharge}
\end{equation}
(this is identical with the sharp-interface prediction of
Eq.~\eqref{eq:SurfaceChargeGC}).

\begin{figure}[tbp]
    \centerline{\includegraphics[width=5in]{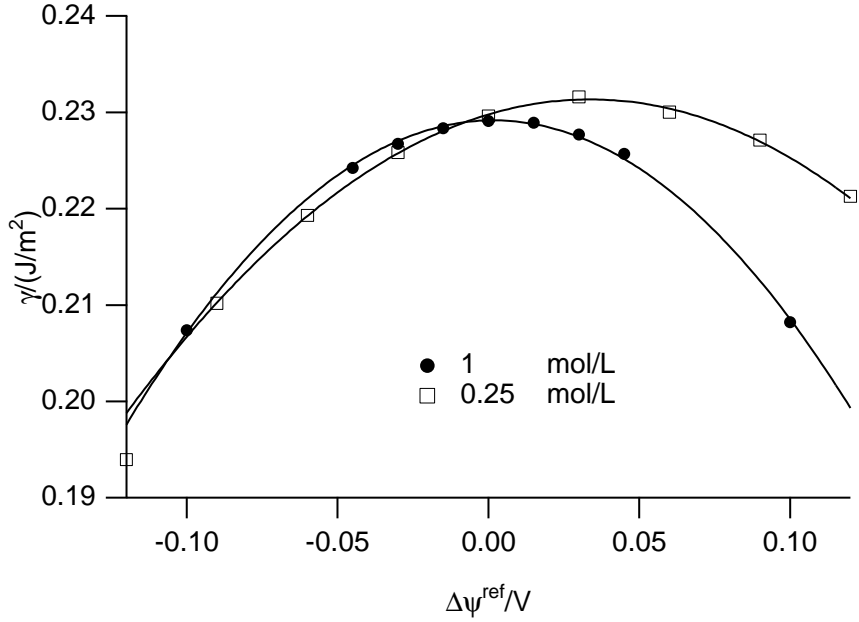}}
    \caption{Surface energy as a function of \Galvani. The lines 
    are to guide the eye.}
    \protect\label{fig:Electrocapillarity}
\end{figure}

\begin{figure}[tbp]
    \centerline{\includegraphics[width=5in]{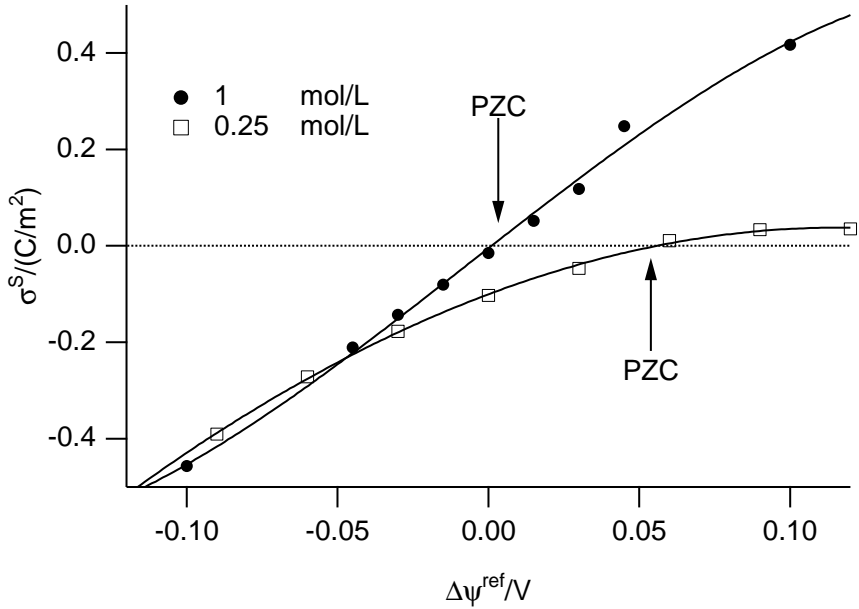}}
    \caption{Electrode surface charge as a function of \Galvani. The lines 
    are to guide the eye.}
    \protect\label{fig:SurfaceCharge}
\end{figure}

We plot the surface energy, calculated with
Eq.~\eqref{eq:SurfaceEnergy}, for different values of \Galvani\ in
Figure~\ref{fig:Electrocapillarity}.
These curves are quite similar to the ``electrocapillary'' curves of
surface energy measured on mercury electrodes \cite{Grahame:1947} for
different applied potentials.  In the electrocapillary experiment, the
applied potential is varied at an inert electrode with a fixed Galvani
potential.  In our simulations, the Galvani potential is varied at a
chemically reactive, but unbiased, interface.  For each curve, there
is a value of \Galvani\ for which the surface energy is a maximum. 
We see from Figure~\ref{fig:SurfaceCharge},
which is calculated from either side of Eq.~\eqref{eq:SurfaceCharge},
that there is a maximum surface energy at the ``potential of zero
charge'', just as is predicted by the Gouy-Chapman model.

\section{Conclusions}

\begin{itemize}
    \item Phase field model of electrochemical interface exhibits
    double layer behavior.
    
    \item Decay length of electrostatic potential are consistent with
    sharp-interface models.

    \item Dependence of surface energy (``electrocapillary curves'')
    and surface charge on potential are consistent with
    sharp-interface models.

    \item Crucial presence of charged species in electrochemistry
    leads to rich interactions between concentration, electrostatic
    potential, and phase stability.
    
    \item Dynamic treatments of this model arise naturally
    \cite{ElPhF:2002}.
\end{itemize}

\section{Acknowledgements}

We are grateful for many fruitful discussions with Ugo Bertocci, Gery
Stafford, Tom Moffat, Daniel Josell, Daniel Wheeler, John Cahn, Sam Coriell,
and Alex Lobkovsky.

\bibliography{abbrTitles,electrochemistry,phaseField}
\bibliographystyle{elsart-num}

\end{document}